\documentclass[journal]{IEEEtran}
\usepackage{stfloats}
\usepackage{float}
\usepackage{graphicx}
\usepackage{subfigure}
\usepackage{color}
\usepackage{algpseudocode}
\usepackage{amssymb}
\usepackage{amsmath}
\usepackage{algorithm}
\usepackage{bm}

\usepackage[colorlinks]{hyperref}

\begin{document}

\title{Exploring Fairness for FAS-assisted Communication Systems: from NOMA  to OMA}
\author{Junteng Yao, Liaoshi Zhou, Tuo Wu, Ming Jin, Cunhua Pan,\\ Maged Elkashlan, and  Kai-Kit Wong, \emph{Fellow}, \emph{IEEE}
\vspace{-5mm}

\thanks{J. Yao, L. Zhou, and M. Jin are with the Faculty of Electrical Engineering and Computer Science, Ningbo University, Ningbo 315211, China (e-mail: $\rm yaojunteng@nbu.edu.cn$, $\rm 2311100202@nbu.edu.cn$, and $\rm jinming@nbu.edu.cn$).}

\thanks{T. Wu  and M. Elkashlan are with the School of Electronic Engineering and Computer Science at Queen Mary University of London, London E1 4NS, U.K. (e-mail: $\rm tuo.wu@qmul.ac.uk$, $\rm maged.elkashlan@qmul.ac.uk$).}

\thanks{C. Pan is with the National Mobile Communications Research Laboratory, Southeast University, Nanjing 210096, China. (e-mail: $\rm cpan@seu.edu.cn$).}

\thanks{K.-K. Wong is with the Department of Electronic and Electrical Engineering, University College London, WC1E 6BT London, U.K., and also with the Yonsei Frontier Laboratory and the School of Integrated Technology, Yonsei University, Seoul 03722, South Korea (e-mail: $\rm kai\text{-}kit.wong@ucl.ac.uk$).}
}

\maketitle

\begin{abstract}
This paper addresses the fairness issue within fluid antenna system (FAS)-assisted non-orthogonal multiple access (NOMA) and orthogonal multiple access (OMA) systems, where a single fixed-antenna base station (BS) transmits superposition-coded signals to two users, each with a single fluid antenna. We define fairness through the minimization of the maximum outage probability for the two users, under total resource constraints for both FAS-assisted NOMA and OMA systems.  Specifically, in the FAS-assisted NOMA systems, we study both a special case and the general case, deriving a closed-form solution for the former and applying a bisection search method to find the optimal solution for the latter. Moreover, for the general case, we derive a locally optimal closed-form solution to achieve fairness. In the FAS-assisted OMA systems, to deal with the non-convex optimization problem with coupling of the variables in the objective function, we employ an approximation strategy to facilitate a successive convex approximation (SCA)-based algorithm, achieving locally optimal solutions for both cases. Empirical analysis validates that our proposed solutions outperform conventional NOMA and OMA benchmarks in terms of fairness.
\end{abstract}


\section{Introduction}
\subsection{Background}
\IEEEPARstart{T}{he advent} of sixth-generation (6G) wireless networks has greatly raised the demand for massive connectivity across a broad spectrum of Internet of Things (IoT) applications, such as industrial automation, environmental monitoring, and autonomous driving, etc. \cite{AGiridhar06,KWChoi18}. The limitations of single-antenna systems have led to the widespread adoption of multiple-input multiple-output (MIMO) in IoT systems to enhance performance \cite{GJ98,LZheng03,JZheng21}. MIMO technology augments communication efficacy by generating spatial bandwidth, independent of frequency and time resources. Nevertheless, MIMO is usually associated with increased energy consumption and hardware costs, attributed to the use of multiple antennas and radio frequency (RF) chains. In response to these challenges, the concept of fluid antenna system (FAS) has recently been introduced. FAS, capable of dynamically changing the antenna position across $N$ prescribed locations (a.k.a.~ports), has been proposed as an effective means to achieve substantial space diversity while minimizing power consumption \cite{KKWong20,KKWong21}.

\subsection{Related Work}
Unlike MIMO which is normally based on multiple fixed-position antennas, a basic FAS can operate with only one RF chain.\footnote{In recent work \cite{New-twc2023}, FAS with multiple RF chains has also been studied.} FAS encompasses any flexible conductive or dielectric structures enabling flexible-position antennas \cite{KKWong221,YHuang21,Zhu-Wong-2024}. Leveraging the position flexibility of antenna, FAS can navigate across a designated area and select the best position for reception. This ability to configure the wireless communication channel has recently garnered significant research interest in single-user \cite{M.Khammassi23,David24,Vega-2023,Vega-2023-2,Alvim-2023,Ghadi-2023,Psomas-dec2023,XLai23,YYe23,BTang23} and multiuser systems \cite{KKWong22,KKWong23,NWaqar23,KKWong231}.

FAS was introduced to wireless communication systems by Wong \emph{et al.} in \cite{KKWong20,KKWong21} where the ergodic capacity and outage probability were analyzed in a point-to-point FAS, where only the receiver was equipped with a fluid antenna. Later in \cite{M.Khammassi23}, Khammassi {\em et al.} proposed an eigenvalue-based channel model to improve the performance analysis of FAS. Most recently, Ramirez-Espinosa {\em et al.} presented a mathematically tractable technique to analyze FAS while ensuring accuracy \cite{David24}. There have been many new results on FAS since, such as \cite{Vega-2023,Vega-2023-2,Alvim-2023,Ghadi-2023,Psomas-dec2023}. Subsequent results further considered the use of maximum ratio combining in FAS with multiple activated ports \cite{XLai23}, the joint optimization of port position and transmit beamforming at the base station (BS) \cite{YYe23}, and extracting the characteristics of FAS for physical layer security \cite{BTang23}.

FAS can also be turned into a multiple access technology. This was first studied in \cite{KKWong22} where it was proposed to switch the position of FAS for the maximization of the instantaneous signal-to-interference plus noise ratio (SINR). Later, a more practical approach, referred to as slow fluid antenna multiple access (FAMA), was proposed in \cite{KKWong23}, where port switching only takes place if the channel changes. Also, deep learning has been employed to facilitate slow FAMA \cite{NWaqar23}. Recently, opportunistic scheduling was also considered in conjunction with FAMA for performance enhancement \cite{KKWong231}. Most recently, a new version of FAMA that could permit spectrum sharing of many users was proposed and analyzed in \cite{Wong-cuma2023}.

It is worth mentioning that channel state information (CSI) is essential at a FAS receiver, like a conventional fixed-position coherent receiver. Great progress has been made towards CSI estimation for FAS, see for example, \cite{HXu23,Dai-2023}.

To manage interference amidst the extensive connectivity in FAS, orthogonal multiple access (OMA) is regarded as a viable solution for serving multiple users. In OMA systems, users are allocated orthogonal resource blocks, such as time, frequency, and code, eliminating the possibility of inter-user interference. Nevertheless, the orthogonal nature of OMA inherently limits the number of accessible users. In addition, channel fading often disrupts this orthogonality \cite{DaiL2015,DingZ2017}. Addressing these limitations, non-orthogonal multiple access (NOMA) has been identified as an efficient strategy for supporting a large number of users within the same physical resource block \cite{DaiL2015,DingZ2017,JYao22}. NOMA operates by transmitting superposition-coded signals, with receivers decoding their intended signals through the successive interference cancellation (SIC) technique, thereby effectively managing interference \cite{LiuYW2023,Zuo2020}.

Research on NOMA categorizes CSI at the transmitter into three types: perfect CSI \cite{YLi18,YSun17,THou20}, partial CSI \cite{THou201,PSwami20,YYapici20}, and statistical CSI \cite{ZDing14,XWang18,SLi20}. Specifically,  Li in \emph{et al.} \cite{YLi18} worked on maximizing the downlink NOMA sum-rate for users distant from the BS by optimizing the transmit beamforming under perfect CSI. Moreover, \cite{YSun17} tackled the weighted sum-rate maximization in multi-carrier NOMA systems, allocating each subcarrier to a maximum of two users in both downlink and uplink. With partial CSI, \cite{THou201} proposed a clustering strategy enhancing the NOMA system coverage. Furthermore, \cite{PSwami20} investigated user scheduling in NOMA with limited feedback, while \cite{YYapici20} assessed the outage performance in millimeter-wave NOMA with one-bit quantized feedback. In the case of statistical CSI, Ding \emph{et al.} derived the outage probability and achievable rate of NOMA using a two dimensional (2D) homogeneous Poisson point process for the BSs and user modeling in \cite{ZDing14}. Afterwards, \cite{XWang18} studied the maximization of the sum-rate of NOMA under Nakagami-$m$ fading channels through a strategic power allocation scheme. Recently, \cite{SLi20} evaluated the outage performance in multi-carrier downlink NOMA systems based on statistical CSI, defining outage as the condition when a user's sum rate across subcarriers falls below a target rate.

The above mentioned works demonstrated the superiority of OMA systems and NOMA systems. Therefore, it is envisioned that enabling the OMA and NOMA techniques in FAS can further improve the quality of services (QoS) \cite{CS23,LT23,JZheng23,WNew23}. Hence, Skouroumounis \emph{et al.} in \cite{CS23} studied the tradeoff between channel estimation and outage probability in the FAS-assisted time-division multiple access (TDMA) systems by using the stochastic geometry method. The authors of \cite{LT23} analyzed the outage performance of the FAS-assisted cooperative NOMA systems, where the central user (CU) serves as a relay for forwarding the signals to the cell-edge users (EUs), while Zheng \emph{et al.}  investigated the outage performance of the FAS-assisted NOMA systems with short-packet communications in \cite{JZheng23}.  The authors of \cite{WNew23} maximized the sum rate in the FAS-assisted NOMA systems by jointly designing port selection and power allocation subject to per-user QoS.

\subsection{Motivations and Contributions}
The aforementioned studies underscored the significant benefits of FAS for both OMA and NOMA  systems. However, these investigations often assumed the availability of perfect instantaneous CSI at the transmitter side, an assumption that is  impractical due to the challenges in acquiring accurate CSI caused by non-ideal feedback channels and delays. Moreover, though the assumption of partial CSI availability presents a more realistic scenario, it significantly complicates the power allocation process in NOMA systems due to the uncertainty in the ordering of channel gains.  Consequently, we adopt a more realistic scenario, namely, statistical CSI. In this context, the outage performance of OMA and NOMA systems emerges as an appropriate metric for analysis and optimization.

In FAS-assisted communication systems, ensuring fairness is crucial for effectively guaranteeing  QoS for different users and reducing the outage performance \cite{CS23,LT23,JZheng23}. The challenge of resource allocation is particularly acute under limited resources. In NOMA, where superposition signals are concurrently transmitted to different users within the same resource blocks, inter-user interference is inevitable. A limited total power budget necessitates that enhancing one user's signal strength comes at the expense of others, leading to higher outage probabilities for users with weakened signal power \cite{RJiao20,SShi16, XZhang19}. Thus, investigating fairness within FAS-assisted NOMA systems is of great importance. Similarly, in FAS-assisted OMA systems, fairness remains critical \cite{XZhang19}.

Motivated by this, this paper aims to tackle the fairness issue in FAS-assisted NOMA and OMA communication systems. The main contributions of this study are outlined below:
\begin{itemize}
\item This paper studies both FAS-assisted NOMA and OMA systems, where the BS is equipped with a single fixed-position antenna, and the CU and the EU are equipped with a single fluid antenna. Assuming that the BS only knows the statistical CSI of the two users, we minimize the maximum of the outage probability of the two users by designing the corresponding resource allocation.
\item In FAS-assisted NOMA systems, we consider a special case and the general case. In the special case, we presume identical parameters across the two users. We prove the existence of a unique optimal closed-form solution. In the general case, deriving a closed-form solution proves challenging due to the integral of the first-order Marcum $Q$-function in the expression of users' outage probabilities. To overcome this, we adopt bisection search, leveraging the monotonic relationship of the outage probabilities with the power allocation coefficient, to find the optimal solution. Additionally, to reduce computational complexity, we then introduce an approximate expression for the objective function and derive a locally optimal solution in closed form for the general case.
\item In FAS-assisted OMA systems, we assume TDMA systems. The problem is non-convex due to coupling of the time and power allocation coefficients in the objective function, rendering the method proposed for NOMA systems ineffective.  Furthermore, the formulation of this objective function presents considerable computational difficulties. To mitigate this challenge, we approximate the objective function and introduce a successive convex approximation (SCA) algorithm to derive a locally optimal solution for both the special and general cases.
\item Our numerical results illustrate that the proposed schemes greatly outperform the benchmarks in conventional OMA and NOMA in terms of outage performance.
\end{itemize}

The rest of this paper is organized as follows. Section \ref{sec2} introduces the network model for the FAS-assisted downlink system. Section \ref{sec3} considers the special and general cases in the case of NOMA and presents the proposed scheme. Section \ref{sec4} then conducts the respective work for the FAS-assisted OMA systems. Simulation results and discussion are provided in Section \ref{sec5}. Finally, conclusions are drawn in Section \ref{sec6}.

\section{FAS Network Model}\label{sec2}
Consider a FAS-assisted NOMA system, which consists of a BS equipped with a single traditional antenna, a CU near the BS, and an EU far from the BS. Both CU and EU are equipped with a single fluid antenna to receive signals from the BS. Both the CU and the EU have the capability to  switch their FASs to the most favourable port among a total of $K$ and $M$ ports, respectively. These ports are evenly distributed across linear spaces of sizes $W_c\lambda$ for the CU and $W_e\lambda$ for the EU, where $\lambda$ represents the wavelength. This arrangement facilitates optimal antenna positioning for both users, enhancing signal reception by leveraging the spatial diversity provided by the multiple ports. Moreover, it is assumed that the time delays resulting from port switching are insignificantly small \cite{KKWong21,KKWong221}.

\subsection{Channel Model}
In the FAS-assisted NOMA system under consideration, the communication channels from the  BS to the $k$-th port of the CU and the $m$-th port of the EU  are represented by $h_k$ and $g_m$, respectively, where $k = 1, \ldots, K$ and $m = 1, \ldots, M$, $K \geq 2$ and $M \geq 2$. The specific mathematical expressions for these channels can be modelled as \cite{KKWong20,KKWong21,KKWong23}
\begin{align}
h_k =& \mu_h h_0 + (1-\mu_h)z_k, \\
g_m=& \mu_g g_0 + (1-\mu_g)\nu_m,
\end{align}
where the channel parameters for the virtual reference ports, $h_0$ for the CU and $g_0$ for the EU, are characterized by complex Gaussian distributions $\mathcal{CN}(0,\sigma_h^2)$ and $\mathcal{CN}(0,\sigma_g^2)$, respectively. Furthermore, the terms $z_k$ and $\nu_j$ are independently and identically distributed (i.i.d.) random variables for the CU and EU, respectively, and follow complex Gaussian distributions with $\mathcal{CN}(0,\sigma_h^2)$ and $\mathcal{CN}(0,\sigma_g^2)$. Additionally, $\mu_h$ and $\mu_g$ are the correlation parameters, given by \cite{KKWong20,KKWong23}
\begin{align}\label{e2}
\mu_h =& \sqrt{2}\sqrt{{}_1F_2\left(\frac{1}{2};1;\frac{3}{2};-\pi^2W_c^2\right)-\frac{J_1(2\pi W_c)}{2\pi W_c}},\\
\mu_g =& \sqrt{2}\sqrt{{}_1F_2\left(\frac{1}{2};1;\frac{3}{2};-\pi^2W_e^2\right)-\frac{J_1(2\pi W_e)}{2\pi W_e}},
\end{align}
where ${}_a F_b$ denotes the generalized hypergeometric function and $J_1(\cdot)$ is the first-order Bessel function of the first kind.

Given $|h_0|$, the probability density function (PDF) of $h_k$ can be expressed as
\begin{align}\label{e3}
f_{|h_k| \big| |h_0|}(r_n | r_0) = \frac{2r_k}{\sigma_h^2 (1-\mu_h^2)} e^{-\frac{r_n^2+\mu_h^2 r_0^2}{\sigma_h^2 (1-\mu_h^2)}} I_{0}\left(\frac{2 \mu_h r_k r_0}{\sigma_h^2 (1-\mu_h^2)}\right),
\end{align}
where $I_{0}(z)$ denotes the zeroth-order modified Bessel function of the first kind and has the following series representation \cite{ISGradshteyn07}
\begin{align}\label{e4}
I_0(z) = \sum_{a=0}^{\infty} \frac{z^{2a}}{2^{2a} a!\Gamma(a+1)},
\end{align}
where $\Gamma(a+1)=a!$.

Similarly, given $|g_0|$, the PDF of $g_m$ is given by
\begin{align}\label{e5}
f_{|g_m| \big| |g_0|}(r_m | r_0) = \frac{2r_m}{\sigma_g^2 (1-\mu_g^2)} e^{-\frac{r_m^2+\mu_g^2 r_0^2}{\sigma_g^2 (1-\mu_g^2)}} I_{0}\left(\frac{2 \mu_g r_m r_0}{\sigma_g^2 (1-\mu_g^2)}\right).
\end{align}

\subsection{NOMA Signal Model}
Building upon the NOMA protocol, the signal transmitted from the BS can be formulated  as
\begin{align}\label{q1}
x = \sqrt{P\alpha}x_c+\sqrt{P(1-\alpha)}x_e,
\end{align}
where $P$ is the transmit power at the BS, and $\alpha$ represents  the power allocation coefficient for the CU. The terms $x_c$ and $x_e$ correspond to the independent, information-bearing signals intended for the CU and EU, respectively. It is assumed that both signals are normalized, i.e., $\mathbf{E}[|x_c|]=\mathbf{E}[|x_e|]=1$.

In this context, let us denote the signals received at the CU and EU as $y_k$ and $y_m$, respectively. These received signals are defined by the following equations:
\begin{align}\label{q2}
y_k&=h_kx+n_k,\\
y_m&=g_mx+n_m,
\end{align}
in which $h_k$ and $g_m$ represent the quasi-static fading channel parameters from the BS to the $k$-th port of the CU and the $m$-th port of the EU, respectively. Furthermore,  $n_k \sim \mathcal{CN}(0,\sigma^2)$ and $n_m \sim \mathcal{CN}(0,\sigma^2)$ denote the independent additive white Gaussian noises at the CU and EU, respectively.

Upon receiving the signal $y_k$, the CU employs the SIC scheme to first decode $x_e$ by regarding $x_c$ as interference. The corresponding SINR for this process is expressed as
\begin{align}\label{q3}
\gamma_{c,e}= \frac{P(1-\alpha)|h_c|^2}{P\alpha|h_c|^2+\sigma^2}.
\end{align}
Given that  $x_e$ is successfully decoded by the CU, the CU then proceeds to decode $x_c$. The signal-to-noise ratio (SNR) for decoding $x_c$ is given by
\begin{align}\label{q4}
\gamma_{c}= \frac{P\alpha|h_c|^2}{\sigma^2}.
\end{align}

Additionally, it is assumed that the CU can swiftly switch its fluid antenna to the most advantageous port, thereby achieving the maximum value of $|h_k|$, hence $|h_c|$. This maximum value can be mathematically expressed as
\begin{align}
|h_{\max}| =\max\{|h_1|, |h_2|, \dots, |h_{K}|\}.
\end{align}
Then the outage probability at the CU is given by
\begin{align}\label{q5}
\mathbf{P}_\textrm{c}^\textrm{out}=& 1-\mathbf{P}(|h_{\max}|^2>\max\{\phi_1,\phi_2\})\nonumber\\
=&\int_0^\infty e^{-t} \cdot \left[1-Q_1\left(\sqrt{\frac{2\mu_h^2}{1-\mu_h^2}}\sqrt{t},\right.\right.\nonumber\\
&\quad\quad\quad\qquad\qquad\left.\left.\sqrt{\frac{2}{1-\mu_h^2}}\sqrt{\frac{\max\{\phi_1,\phi_2\}}
{\sigma_h^2}}\right)\right]^{K}dt,
\end{align}
where
\begin{align}
\label{q6}\phi_1= &\frac{\gamma_{th,2}\sigma^2}{P(1-\alpha-\gamma_{th,2}\alpha)},\\
\label{q7}\phi_2= &\frac{\gamma_{th,1}\sigma^2}{P\alpha},
\end{align}
$\gamma_{th,1}=2^{R_1}-1$ and $\gamma_{th,2}=2^{R_2}-1$, $R_1$ and $R_2$ are the targeted rate of $x_c$ and $x_e$, respectively.

From \eqref{q6}, we know that $\phi_1>0$ when $\alpha<\frac{1}{1+\gamma_{th,2}}$, and $\mathbf{P}_c^\textrm{out}=1$ when $\phi_1\leq0$. Combining \eqref{q6} with \eqref{q7}, we know that $\phi_1> \phi_2$ when $\frac{1}{1+\gamma_{th,2}+\gamma_{th,2}/\gamma_{th,1}}< \alpha<\frac{1}{1+\gamma_{th,2}}$, $\phi_1\leq \phi_2$ when $0<\alpha\leq \frac{1}{1+\gamma_{th,2}+\gamma_{th,2}/\gamma_{th,1}}$.

For the EU, after receiving $y_m$, it decodes $x_e$ by treating $x_c$ as interference, and the corresponding SINR is given by
\begin{align}\label{q8}
\gamma_{e}^m= \frac{P(1-\alpha)|g_m|^2}{P\alpha|g_m|^2+\sigma^2}.
\end{align}
Additionally, it is assumed that the EU can swiftly switch its fluid antenna to the optimal port. Consequently, the maximum value of $|g_m|$ can be expressed as
\begin{align}
|g_{\max}| =\max\{|g_1|, |g_2|, \dots, |g_{M}|\}.
\end{align}
In this case, the outage probability at EU is expressed as
\begin{align}\label{q9}
\mathbf{P}_e^\textrm{out}= &1-\mathbf{P}(|g_{\max}|^2>\phi_1)\nonumber\\
=&\int_0^\infty e^{-t} \left[1-Q_1\left(\sqrt{\frac{2\mu_g^2}{1-\mu_g^2}}\sqrt{t}, \right.\right.\nonumber\\
&\quad\quad\quad\qquad\qquad\left.\left.
\sqrt{\frac{2}{1-\mu_g^2}}\sqrt{\frac{\phi_1}{\sigma_g^2}}\right)\right]^{M}dt.
\end{align}

\subsection{Problem Formulation}
To enhance the spectrum efficiency of the NOMA system, addressing the fairness issue is crucial when optimizing the overall communication performance of the systems. To impose fairness, here, we consider the min-max outage probability optimization problem, which can be formulated as
\begin{align}
\label{q10} &\min_{\alpha} \ \max\{\mathbf{P}_c^\textrm{out},\mathbf{P}_e^\textrm{out}\} \notag\\
&\ \ \mbox{s.t.}\ 0<\alpha<\frac{1}{1+\gamma_{th,2}}.
\end{align}
According to \eqref{q6} and \eqref{q7}, Problem \eqref{q10} can be decomposed into the following two sub-problems:
\begin{align}\label{q11}
\textrm{(P1)} \quad \min_{\alpha}& \ \max\{\mathbf{P}_c^\textrm{out}(\phi_2),\mathbf{P}_e^\textrm{out}\} \nonumber\\
\mbox{s.t.}&\ 0<\alpha\leq \frac{1}{1+\gamma_{th,2}+\gamma_{th,2}/\gamma_{th,1}},
\end{align}
and
\begin{align}\label{q12}
\textrm{(P2)} \quad \min_{\alpha}& \ \max\{\mathbf{P}_c^\textrm{out}(\phi_1),\mathbf{P}_e^\textrm{out}\} \nonumber\\
\mbox{s.t.}&\ \frac{1}{1+\gamma_{th,2}+\gamma_{th,2}/\gamma_{th,1}}< \alpha<\frac{1}{1+\gamma_{th,2}},
\end{align}
where
\begin{align}
\label{q13}\mathbf{P}_c^\textrm{out}(\phi_1)=&\int_0^\infty e^{-t} \left[1-Q_1\left(\sqrt{\frac{2\mu_h^2}{1-\mu_h^2}}\sqrt{t},\right.\right.\nonumber\\
&\quad\quad\quad\qquad\qquad\left.\left.\sqrt{\frac{2}{1-\mu_h^2}}\sqrt{\frac{\phi_1}{\sigma_h^2}}\right)\right]^{K}dt,\\
\label{q14}\mathbf{P}_c^\textrm{out}(\phi_2)=&\int_0^\infty e^{-t} \left[1-Q_1\left(\sqrt{\frac{2\mu_h^2}{1-\mu_h^2}}\sqrt{t},\right.\right.\nonumber\\
&\quad\quad\quad\qquad\qquad\left.\left.\sqrt{\frac{2}{1-\mu_h^2}}\sqrt{\frac{\phi_2}{\sigma_h^2}}\right)\right]^{M}dt.
\end{align}

\section{Min-Max Fairness for NOMA Systems}\label{sec3}
In this section, we aim to solve Problem \eqref{q10} in two different cases according to the situation of the FAS within the CU and that within the EU. In what follows, we then deduce several insights for the considered FAS-assisted NOMA systems.

\subsection{Special Case}
In this subsection, we examine a particular case where $K=M=N$ and $\mu_h=\mu_g=\mu$. Under this condition, we first investigate the monotonicity of $\mathbf{P}_c^\textrm{out}(\phi_1)$, $\mathbf{P}_c^\textrm{out}(\phi_2)$, and $\mathbf{P}_e^\textrm{out}$ with respect to $\alpha$, as detailed in the subsequent lemma.

\emph{Lemma 1}: $\mathbf{P}_c^\textrm{out}(\phi_2)$ decreases with $\alpha$, whereas $\mathbf{P}_c^\textrm{out}(\phi_1)$ and $\mathbf{P}_e^\textrm{out}$ increase with $\alpha$.

\emph{Proof}: See Appendix A. $\hfill\blacksquare$

Leveraging the monotonicity established in \emph{Lemma 1}, the following lemma addresses Problem \eqref{q10}.

\emph{Lemma 2}: The optimal solution for Problem \eqref{q10} is
\begin{equation}
\frac{1}{1+\gamma_{th,2}+\frac{\gamma_{th,2}\sigma_h^2}{\gamma_{th,1}\sigma_g^2}}.
\end{equation}

\emph{Proof}: See Appendix B. $\hfill\blacksquare$

\emph{Remark 1}: The optimal solution identified in \eqref{bq1} for this specific case aligns with that of traditional NOMA using a conventional single antenna. Thus, when the number of ports for the CU and EU are equal, the optimization algorithm from traditional NOMA approaches can be applied.

\subsection{General Case}
In this subsection, the analysis extends to a general scenario where $N_e \neq N_c$ or $\mu_h \neq \mu_g$, aiming to solve Problem \eqref{q10}.

Drawing from \emph{Lemma 1} with the special case, the monotonic behaviors of $\mathbf{P}_c^\textrm{out}(\phi_2)$, $\mathbf{P}_c^\textrm{out}(\phi_1)$, and $\mathbf{P}_e^\textrm{out}$ in relation to $\alpha$ are consistent in the general case. Consequently, given that the solution to Problem \eqref{q10} coincides with that of Problem \eqref{q11}, we deduce that these problems are effectively synonymous.

However, given the differences $N_e \neq N_c$ and $\mu_h \neq \mu_g$, we cannot assert that $\mathbf{P}_c^\textrm{out}(\phi_2) = \mathbf{P}_c^\textrm{out}(\phi_1) < \mathbf{P}_e^\textrm{out}$ under the condition $\alpha = \frac{1}{1+\gamma{th,2}+\gamma{th,2}/\gamma{th,1}}$, as proposed in the special case. Notably, when $\alpha = 0$, $\mathbf{P}_e^\textrm{out} < \mathbf{P}_c^\textrm{out}(\phi_2) = 1$, leading to the identification of optimal solutions for Problem \eqref{q11} in the following two conditions:

\emph{Condition 1}: The optimal $\alpha$ is $\frac{1}{1+\gamma_{th,2}+\gamma_{th,2}/\gamma_{th,1}}$ if $\mathbf{P}_c^\textrm{out}(\phi_2) \geq \mathbf{P}_e^\textrm{out}$ at this $\alpha$ value;

\emph{Condition 2}:  The optimal $\alpha$ lies within $\left(0, \frac{1}{1+\gamma_{th,2}+\gamma_{th,2}/\gamma_{th,1}}\right)$ if $\mathbf{P}_c^\textrm{out}(\phi_2) < \mathbf{P}_e^\textrm{out}$ at the specified $\alpha$.

For \emph{Condition 2}, a unique value of $\alpha$ exists that equates $\mathbf{P}_c^\textrm{out}(\phi_2)$ with $\mathbf{P}_e^\textrm{out}$. To determine the optimal $\alpha$ under \emph{Condition 2}, the bisection search method is employed within the interval $\left(0, \frac{1}{1+\gamma_{th,2}+\gamma_{th,2}/\gamma_{th,1}}\right)$.

\emph{Complexity Analysis}: The complexity of solving \eqref{q11} is
\begin{equation}
\mathcal{O}\left(\log \frac{1}{(1+\gamma_{th,2}+\gamma_{th,2}/\gamma_{th,1})\delta} \right),
\end{equation}
where $\delta$ is the accuracy of the bisection search.

The complexity analysis above reveals that the computational complexity of utilizing the bisection search method is considerable. To further reduce computational complexity, we introduce a method based on a closed-form solution to address Problem \eqref{q11}. Specifically, for $\mathbf{P}^\textrm{out}_{c}(\phi_2)$, given that $0 \leq Q_1\left(\sqrt{\frac{2\mu_h^2}{1-\mu_h^2}}\sqrt{t},\sqrt{\frac{2}{1-\mu_h^2}}\sqrt{\frac{\phi_2}{\sigma_h^2}}\right) \leq 1$, the following approximation can be derived:
\begin{multline}\label{qq36}
\left[1-Q_1\left(\sqrt{\frac{2\mu_h^2}{1-\mu_h^2}}\sqrt{t},\sqrt{\frac{2}{1-\mu_h^2}}\sqrt{\frac{\phi_2}{\sigma_h^2}}\right)\right]^{N_c}\\
\approx 1-N_cQ_1\left(\sqrt{\frac{2\mu_h^2}{1-\mu_h^2}}\sqrt{t},\sqrt{\frac{2}{1-\mu_h^2}}\sqrt{\frac{\phi_2}{\sigma_h^2}}\right).
\end{multline}
Consequently, by using
\begin{multline}\label{q27}
\int_0^c e^{-t}Q_1(a\sqrt{t},b)dt=e^{-\frac{b^2}{a^2+2}}Q_1\left(\sqrt{c(a^2+2)},\frac{ab}{\sqrt{a^2+2}}\right)\\
-e^{-c}Q_1(a\sqrt{c},b),
\end{multline}
$\mathbf{P}^\textrm{out}_{c}(\phi_2)$ can be approximated by
\begin{align}\label{q28}
\tilde{\mathbf{P}}^\textrm{out}_{c}(\phi_2)=&1-N_ce^{-\frac{\phi_2}{\sigma_h^2}}.
\end{align}

Similarly, $\mathbf{P}^\textrm{out}_{e}$ can be approximated as
\begin{align}\label{q29}
\tilde{\mathbf{P}}^\textrm{out}_{e} = &1 - N_e e^{-\frac{\phi_1}{\sigma_g^2}}.
\end{align}
Drawing parallels to \emph{Lemma }1, it becomes apparent that $\tilde{\mathbf{P}}^\textrm{out}_{c}(\phi_2)$ decreases with $\alpha$, whereas $\tilde{\mathbf{P}}^\textrm{out}_{e}$ increases with $\alpha$. Consequently, the optimal solution to Problem \eqref{q11} is achieved when $\tilde{\mathbf{P}}^\textrm{out}_{c}(\phi_2) = \tilde{\mathbf{P}}^\textrm{out}_{e}$. This leads  to the following theorem.

\emph{Theorem 1}: For the solution to Problem \eqref{q11},
\begin{itemize}
    \item When $N_c = N_e$, the closed-form solution is given by
    \begin{align}\label{q31}
    \tilde{\alpha} = & \frac{1}{1 + \gamma_{th,2} + \frac{\gamma_{th,2} \sigma_h^2}{\gamma_{th,1} \sigma_g^2}}.
    \end{align}
    \item When $N_c \neq N_e$, and both $\alpha_1 \geq 0$ and $\alpha_2 \geq 0$ are real values, the closed-form solution is found as
    \begin{align}\label{q30}
    \tilde{\alpha} = & \min\left\{ \min\left\{ [\alpha_1]^+, \frac{1}{1 + \gamma_{th,2} + \frac{\gamma_{th,2}}{\gamma_{th,1}}} \right\}, \right.\nonumber\\
    & \left. \min\left\{ [\alpha_2]^+, \frac{1}{1 + \gamma_{th,2} + \frac{\gamma_{th,2}}{\gamma_{th,1}}} \right\} \right\},
    \end{align}
\end{itemize}
where $[z]^+$ denotes $\max\{z, 0\}$, and $\alpha_1$ and $\alpha_2$ are determined by \eqref{eq5} and \eqref{eq6}, respectively.

\emph{Proof}: See Appendix C. $\hfill\blacksquare$

If $\alpha_1$ or $\alpha_2$ are not real values, they are set to $+\infty$. Similarly, if $\alpha_1 < 0$ or $\alpha_2 < 0$, they are also set to $+\infty$. The optimal solution is thereby determined by \eqref{q30}.

\emph{Remark 2}: When $N_c = N_e$, the optimal solution given by \eqref{q31} aligns with the solution in \eqref{bq1} for the special case.

\section{Min-Max Fairness for OMA systems}\label{sec4}
In this section, we extend our investigation to conventional OMA  systems, i.e.,  TDMA, to provide a comprehensive understanding of fairness issues within FAS-assisted communication systems. To be specific, we delve into the fairness optimization problem in OMA systems, which serves as a benchmark for comparing these systems against NOMA systems.

\subsection{Problem Formulation}
Based on the TDMA protocol, the  SNR at the  CU and the EU can be, respectively, expressed as
\begin{align}\label{q32}
\gamma_{c,\textrm{OMA}}^k = & \frac{P\alpha |h_k|^2}{\sigma^2}, \\
\gamma_{e,\textrm{OMA}}^m = & \frac{P(1-\alpha) |g_m|^2}{\sigma^2}.
\end{align}
Then by denoting the normalized time allocation coefficient for the CU as $\beta$ (where $0 < \beta < 1$), the achievable rates of CU and EU can be derived as
\begin{align}\label{q33}
R_c = & \, \beta \log_2\left(1 + \max\{\gamma_{c,\textrm{OMA}}^k\}\right), \\
R_e = & \, (1 - \beta) \log_2\left(1 + \max\{\gamma_{e,\textrm{OMA}}^m\}\right).
\end{align}
Accordingly, the outage probability at the  CU and the EU can be written as
\begin{align}\label{q34}
\mathbf{P}_{c,\textrm{OMA}}^\textrm{out}=\int_0^\infty e^{-t} \cdot & \left[1-Q_1\left(\sqrt{\frac{2\mu_h^2}{1-\mu_h^2}}\sqrt{t},\right.\right.\nonumber\\
&\left.\left.\sqrt{\frac{2}{1-\mu_h^2}}\sqrt{\frac{\psi_1}{\sigma_h^2}}\right)\right]^{N_c}dt,\\
\mathbf{P}_{e,\textrm{OMA}}^\textrm{out}=\int_0^\infty e^{-t} \cdot & \left[1-Q_1\left(\sqrt{\frac{2\mu_g^2}{1-\mu_g^2}}\sqrt{t},\right.\right.\nonumber\\
&\left.\left.\sqrt{\frac{2}{1-\mu_g^2}}\sqrt{\frac{\psi_2}{\sigma_g^2}}\right)\right]^{N_e}dt,
\end{align}
where
$\psi_1= \frac{\bar{\gamma}_{th,1}\sigma^2}{P\alpha}$, $\psi_2= \frac{\bar{\gamma}_{th,2}\sigma^2}{P(1-\alpha)}$, $\bar{\gamma}_{th,1}=2^{\frac{R_1}{\beta}}-1$ and $\bar{\gamma}_{th,2}=2^{\frac{R_2}{1-\beta}}-1$.
Therefore, the min-max fairness optimization problem in OMA systems can be formulated as
\begin{align}\label{q37}
\min_{0<\alpha<1,0<\beta<1}& \ \max\{\mathbf{P}_{c,\textrm{OMA}}^\textrm{out}, \mathbf{P}_{e, \textrm{OMA}}^\textrm{out}\}.
\end{align}

\subsection{SCA-Based Algorithm}
Due to coupling of the optimization variables in the objective function, \eqref{q37} is non-convex. The interdependence of $\alpha$ and $\beta$ in $\mathbf{P}_{c, \textrm{OMA}}^\textrm{out}$ and $\mathbf{P}_{e, \textrm{OMA}}^\textrm{out}$ renders the closed-form solution for the special case and the bisection search for the general case inapplicable to OMA systems. To solve \eqref{q37}, we introduce an SCA-based algorithm to obtain a locally optimal solution.

We begin by deriving the approximations for $\mathbf{P}_c^\text{out}$ and $\mathbf{P}_e^\text{out}$. Following \eqref{e5}, the approximation for $f_{\lvert g_m \rvert \big|  \lvert g_0 \rvert}(r_m \lvert r_0)$ is
\begin{align}\label{q15}
\hat{f}_{\lvert g_m \rvert \big|  \lvert g_0 \rvert}(r_m \lvert r_0) = & \frac{2r_m}{\sigma_g^2 (1-\mu_g^2)} e^{-\frac{r_m^2 + \mu_g^2 r_0^2}{\sigma_g^2 (1-\mu_g^2)}},
\end{align}
when  $I_0(z) \approx 1$.

Considering $|g_0|$, it is evident that $|g_1|, \ldots, |g_{M}|$ are mutually independent. This independence allows us to deduce the approximate joint PDF of $|g_1|, \ldots, |g_{M}|$ given $|g_0|$ as
\begin{align}\label{q16}
\hat{f}_{|g_1|, \ldots, |g_{N_e}| \mid |g_0|}(r_1, \ldots, r_M \mid r_0) = & \prod_{k=1}^{M}\frac{2r_k}{\sigma_g^2 (1-\mu_g^2)} e^{-\frac{r_k^2 + \mu_g^2 r_0^2}{\sigma_g^2 (1-\mu_g^2)}}.
\end{align}
Consequently, the approximate joint PDF of $|g_0|, \ldots, |g_{M}|$ can be expressed as
\begin{align}\label{q17}
&\hat{f}_{|g_0|, \ldots, |g_{M}|}(r_0, \ldots, r_M)\nonumber\\
 &=  \frac{2r_0}{\sigma_g^2} e^{-\frac{r_0^2}{\sigma_g^2}}\prod_{m=1}^{M}\frac{2r_m}{\sigma_g^2 (1-\mu_g^2)} e^{-\frac{r_m^2 + \mu_g^2 r_0^2}{\sigma_g^2 (1-\mu_g^2)}}.
\end{align}
Building upon \eqref{q17}, the approximations of $\mathbf{P}_{c, \text{OMA}}^\text{out}$ and $\mathbf{P}_{e, \text{OMA}}^\text{out}$ are given by
\begin{align}
\label{qq38}\hat{\mathbf{P}}_{c, \text{OMA}}^\text{out} &\approx  \eta_h \left[1 - e^{-\frac{\psi_1}{\sigma_h^2(1-\mu_h^2)}}\right]^{N_c}, \\
\label{qq39}\hat{\mathbf{P}}_{e, \text{OMA}}^\text{out} &\approx  \eta_g \left[1 - e^{-\frac{\psi_2}{\sigma_g^2(1-\mu_g^2)}}\right]^{N_e},
\end{align}
where $\eta_h = \frac{1-\mu_h^2}{1+(N_c-1)\mu_h^2}$, and $\eta_g = \frac{1-\mu_g^2}{1+(N_e-1)\mu_g^2}$.

By introducing a slack variable $\tau$ and leveraging the approximations in \eqref{qq38} and \eqref{qq39}, \eqref{q37} can be rewritten as
\begin{subequations}\label{q38}
\begin{align}
\min_{ \alpha, 0<\rho} &\ \ \tau \\
\mbox{s.t.} &\ \ \hat{\mathbf{P}}_{c, \textrm{OMA}}^\textrm{out}\leq \tau, \label{q38b}\\
&\ \ \hat{\mathbf{P}}_{e, \textrm{OMA}}^\textrm{out}\leq \tau, \label{q38c}\\
&\ \  0<\alpha<1,\\
&\ \ 0<\beta<1.
\end{align}
\end{subequations}
However, \eqref{q38} is still a non-convex problem due to the non-convex constraints \eqref{q38b} and \eqref{q38c}. To tackle this, we further introduce two slack variables, i.e., $a$ and $b$. As such, constraints \eqref{q38b} and \eqref{q38c} can be reformulated as
\begin{align}
\label{q22-1}1-a \leq \left(\frac{\tau}{\eta_h}\right)^{\frac{1}{N_c}} \ \mathrm{and} \  &1-b \leq \left(\frac{\tau}{\eta_g}\right)^{\frac{1}{N_e}}, \\
\label{q22-2}a\leq e^{-\frac{\psi_1}{\sigma_h^2(1-\mu_h^2)}} \ \mathrm{and} \ &b\leq e^{-\frac{\psi_2}{\sigma_g^2(1-\mu_g^2)}}.
\end{align}

For constraints \eqref{q22-1}, we have the following lemma.

\emph{Lemma 3:} The constraints in \eqref{q22-1} are convex.

\emph{Proof:} By taking the first-order partial derivative and the second-order partial derivative of $\left(\frac{\tau}{\eta_g}\right)^{\frac{1}{N_e}}$ with respect to $\tau$, we have
\begin{align}
\label{q23-1}\frac{\partial \left(\frac{\tau}{\eta_g}\right)^{\frac{1}{N_e}}}{\partial \tau}=&\frac{1}{\eta_gN_e}\left(\frac{\tau}{\eta_g}\right)^{\frac{1-N_e}{N_e}}, \\
\label{q23-2}\frac{\partial^2 \left(\frac{\tau}{\eta_g}\right)^{\frac{1}{N_e}}}{\partial \tau^2}=&\frac{1-N_e}{\eta^2_gN^2_e}\left(\frac{\tau}{\eta_g}\right)^{\frac{1-2N_e}{N_e}}.
\end{align}
Because $N_e>1$, $\frac{\partial^2 \left(\frac{\tau}{\eta_g}\right)^{\frac{1}{N_e}}}{\partial \tau^2}<0$. $\left(\frac{\tau}{\eta_g}\right)^{\frac{1}{N_e}}$ is a concave function with respect to $\tau$. Similarly, $\left(\frac{\tau}{\eta_h}\right)^{\frac{1}{N_c}}$ is also a concave function with respect to $\tau$. Therefore, the constraints in \eqref{q22-1} are convex, which completes the proof. $\hfill\blacksquare$

For  constraints \eqref{q22-2}, we introduce two slack variables $c$ and $d$ to address their non-convex property, resulting in the following new constraints:
\begin{align}
\label{q39-1}\ln(a) + \frac{c}{\sigma_h^2(1-\mu_h^2)} \leq 0 & \ \mathrm{and} \ \ln(b) + \frac{d}{\sigma_g^2(1-\mu_g^2)} \leq 0,\\
\label{q39-2}\psi_1 \leq c & \ \mathrm{and} \ \psi_2 \leq d.
\end{align}
Given that the constraints in \eqref{q39-1} include logarithmic functions, we employ the SCA-based algorithm to resolve this.

First, let us define
\begin{align}
\label{q23}\varphi(a) = & \ln(a),\\
\label{q24}\eta(b) = & \ln(b),
\end{align}
and the first-order Taylor expansions around the points $\tilde{a}$ and $\tilde{b}$ are given as
\begin{align}
\label{q24-1}\varphi(a;\tilde{a}) = & \varphi(\tilde{a}) + \frac{a - \tilde{a}}{\tilde{a}},\\
\label{q24-2}\eta(b;\tilde{b}) = & \eta(\tilde{b}) + \frac{b - \tilde{b}}{\tilde{b}},
\end{align}
where $\varphi(a) \leq \varphi(a;\tilde{a})$ and $\eta(b) \leq \eta(b;\tilde{b})$. Thus, constraints \eqref{q39-1} can be reformulated as
\begin{align}
\label{q40}\varphi(a;\tilde{a}) + \frac{c}{\sigma_h^2(1-\mu_h^2)} \leq 0,\\
\label{q41}\eta(b;\tilde{b}) + \frac{d}{\sigma_g^2(1-\mu_g^2)} \leq 0.
\end{align}
Furthermore, the constraints $\psi_1 \leq c$ and $\psi_2 \leq d$ in \eqref{q39-2} can be rewritten as
\begin{align}
\label{qq42}2^{\frac{R_1}{\beta}} - 1 \leq & \frac{P}{\sigma^2}\alpha c,\\
\label{qq43}2^{\frac{R_2}{1-\beta}} - 1 \leq & \frac{P}{\sigma^2}(d - \alpha d).
\end{align}

For constraints \eqref{qq42} and \eqref{qq43}, we have following lemma.

\emph{Lemma 4}: $2^{\frac{R_1}{\beta}}$ and  $2^{\frac{R_2}{1-\beta}}$ are both convex functions with respect to $\beta$.

\emph{Proof}: See Appendix D. $\hfill\blacksquare$

We also ascertain that
\begin{align}\label{q43}
\alpha c=\frac{1}{4}(\alpha+c)^2-\frac{1}{4}(\alpha-c)^2.
\end{align}
Owing to the fact that $\frac{1}{4}(\alpha + c)^2$ and $\frac{1}{4}(\alpha - c)^2$ are both convex with respect to $\alpha$ and $c$. The first-order Taylor expansion of $\frac{1}{4}(\alpha + c)^2$ around the point $(\tilde{\alpha}, \tilde{c})$ is computed as
\begin{align}\label{q44}
\zeta(\alpha, c;\tilde{\alpha}, \tilde{c}) &= \frac{1}{4}(\tilde{\alpha}+\tilde{c})^2+\frac{1}{2}(\tilde{\alpha}+\tilde{c})(\alpha-\tilde{\alpha}+c-\tilde{c}).
\end{align}

Thus, constraint \eqref{qq42} can be recast as
\begin{align}\label{q45}
2^{\frac{R_1}{\beta}}-1 \leq& \frac{P}{\sigma^2}\left(\zeta(\alpha, c;\tilde{\alpha}, \tilde{c}) -\frac{1}{4}(\alpha-c)^2\right).
\end{align}

Similar to \eqref{q44}, the first-order Taylor expansion of $\frac{1}{4}(\alpha-d)^2$ around the point ($\tilde{\alpha}, \tilde{d}$) is calculated as
\begin{align}\label{q46}
\xi(\alpha, d;\tilde{\alpha}, \tilde{d}) &= \frac{1}{4}(\tilde{\alpha}-\tilde{d})^2+\frac{1}{2}(\tilde{\alpha}-\tilde{d})(\alpha-\tilde{\alpha}-d+\tilde{d}).
\end{align}
As a result, constraint \eqref{qq43} can be rewritten as
\begin{align}\label{q47}
2^{\frac{R_2}{1-\beta}}-1 \leq& \frac{P}{\sigma^2}\left(d-\frac{1}{4}(\alpha+d)^2+\xi(\alpha, d;\tilde{\alpha}, \tilde{d}) \right).
\end{align}

In the $(m+1)$-th iteration, given $\alpha^{(m)}$, $a^{(m)}$, $b^{(m)}$, $c^{(m)}$, and $d^{(m)}$ as the optimal solutions for $\alpha$, $a$, $b$, $c$, and $d$ from the $m$-th iteration, we present the following problem:
\begin{subequations}\label{q48}
\begin{align}
\min_{\Xi} \ & \tau \\
\text{s.t.} \
& \varphi(a;a^{(m)})+\frac{c}{\sigma_h^2(1-\mu_h^2)} \leq 0, \label{q48b} \\
& \eta(b;b^{(m)})+\frac{d}{\sigma_g^2(1-\mu_g^2)} \leq 0, \label{q48c} \\
& 2^{\frac{R_1}{\beta}}-1 \leq \frac{P}{\sigma^2}\left(\zeta(\alpha, c;\alpha^{(m)}, c^{(m)}) - \frac{1}{4}(\alpha-c)^2\right), \\
& 2^{\frac{R_2}{1-\beta}}-1 \leq \frac{P}{\sigma^2}\left(d-\frac{1}{4}(\alpha+d)^2+\xi(\alpha, d;\alpha^{(m)}, d^{(m)})\right), \\
& \eqref{q22-1},
\end{align}
\end{subequations}
where $\Xi = \{\tau > 0, 1 > \alpha > 0, 1 > \beta > 0, a > 0, b > 0, c > 0, d > 0\}$ and \eqref{q48} is convex and can be solved by CVX.

\emph{Complexity Analysis}: The complexity of updating $\alpha$, $\beta$, $\tau$, $a$, $b$, $c$, and $d$ by using  the interior-point method is $\mathcal{O}\left(\log \frac{1}{\delta}\right)$ \cite{Boyd,IPolik10}, where $\epsilon$ is the accuracy. Thus, the complexity of the SCA based algorithm is found as
\begin{equation}
\mathcal{O}\left(7L\log \frac{1}{\epsilon} \right),
\end{equation}
where $L$ is the number of iterations for convergence.

\section{Numerical Results}\label{sec5}
In our simulation experiments, we assume $\sigma_i^2 = d_i^{-\theta}$ for $i \in \{h, g\}$, where $\theta$ represents the path loss factor, and $d_h$ and $d_g$ denote the distances between the BS and the CU, and the distance between the BS and the EU, respectively. Specifically, we set $d_h = 400$ m, $d_g = 600$ m, and $\theta = 3$. The power of additive Gaussian noises is set by $\sigma^2 = -80$ dBm.

In Fig.~1, we examine the impact of the power allocation coefficient for the CU, denoted as $\alpha$, on the outage probabilities for both CU and EU in the special case of NOMA systems, where $N_c = N_e = 4$, $W_c = W_e = 5$, $R_1 = R_2 = 1$ bps/Hz, and $P = 5$ dBm. Observations from the results in Fig.~1 reveal that the optimal solution is $\frac{1}{1+\gamma_{th,2}+\frac{\gamma_{th,2}\sigma_h^2}{\gamma_{th,1}\sigma_g^2}} = 0.186$, and $\mathbf{P}_c^\textrm{out}(\phi_2) = \mathbf{P}_c^\textrm{out}(\phi_1)$ when $\alpha = \frac{1}{1+\gamma_{th,2}+\gamma_{th,2}/\gamma_{th,1}} = 0.3333$. These findings agree with the analysis in Lemma 1.

\begin{figure}
\centering
\includegraphics[width=3.4in]{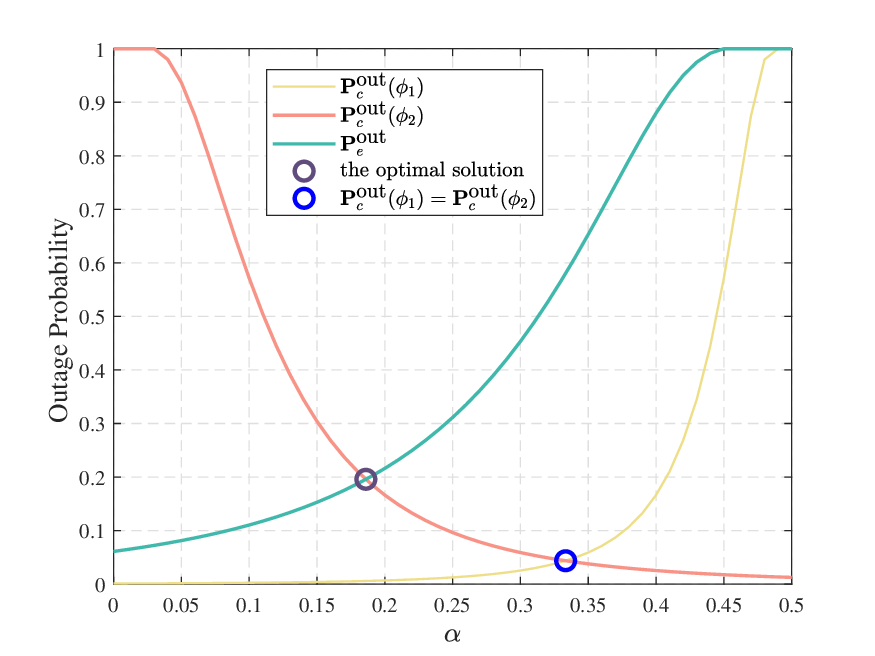}
\caption{Outage probability versus $\alpha$ for the special case in NOMA systems, where $N_c=N_e=4$, $W_c=W_e=5$, $R_1=R_2=1$ bps/Hz, and $P=5$ dBm.}
\end{figure}

In Figs.~2 and 3, we study the impact of the power allocation coefficient $\alpha$ for the CU on the outage probabilities of both CU and EU in NOMA systems under the general case. Analysis of Figs.~2 and 3 reveals two scenarios for optimal solutions when $N_e \neq N_c$ and $\mu_h \neq \mu_g$. Specifically, Fig. 2 illustrates the first scenario, while Fig. 3 depicts the second. In Fig. 2, the optimal solution is $\alpha = \frac{1}{1 + \gamma_{th,2} + \gamma_{th,2}/\gamma_{th,1}} = 0.3333$, and in Fig. 3, a bisection search yields an optimal solution of $0.2887$.

\begin{figure}
\centering
\includegraphics[width=3.4in]{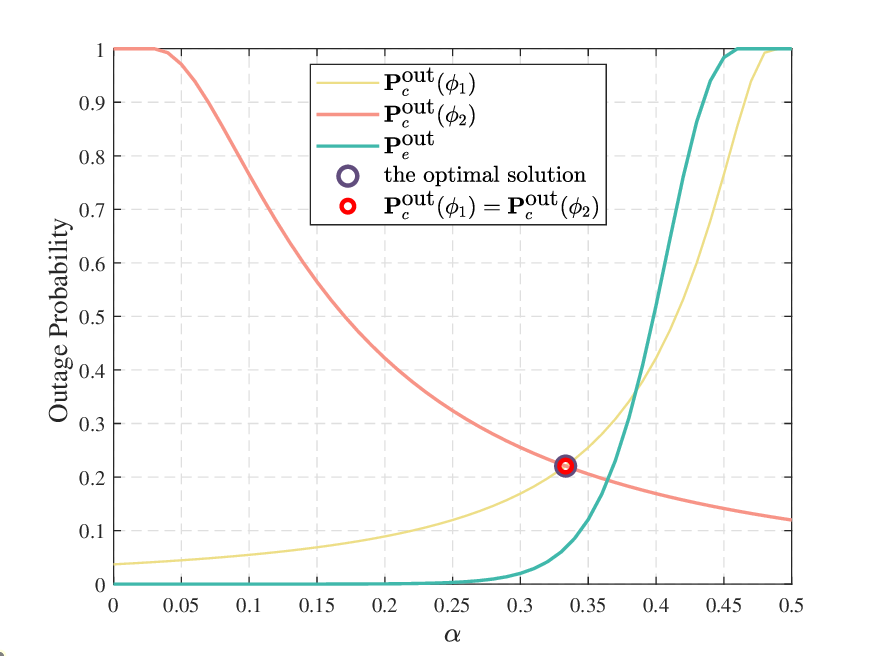}
\caption{Outage probability versus $\alpha$ for case 1 of the general case in NOMA systems, where $N_c=2$, $N_e=20$, $W_c=1$, $W_e=5$, $R_1=R_2=1$ bps/Hz, and $P=5$ dBm.}
\end{figure}

\begin{figure}
\centering
\includegraphics[width=3.4in]{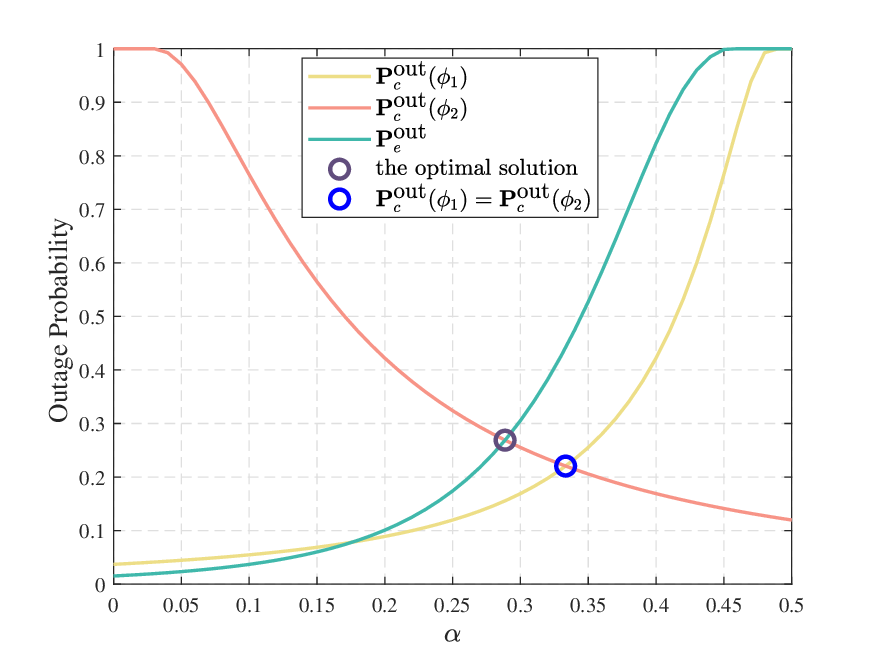}
\caption{Outage probability versus $\alpha$ for case 2 of the general case in NOMA systems, where $N_c=2$, $N_e=6$, $W_c=1$, $W_e=5$, $R_1=R_2=1$ bps/Hz, and $P=5$ dBm.}
\end{figure}

Fig.~4 illustrates the convergence behavior of our proposed SCA-based algorithm for the considered OMA systems, characterized by equal numbers of CUs and EUs $(N_c=N_e=4)$, bandwidths $(W_c=W_e=5)$, rate requirements ($R_1=R_2=1$ bps/Hz), and transmission powers ($P=5$ dBm, $10$ dBm, and $15$ dBm). The algorithm demonstrates rapid convergence, typically within $2$ iterations, as depicted in Fig.~4.

\begin{figure}
\centering
\includegraphics[width=3.4in]{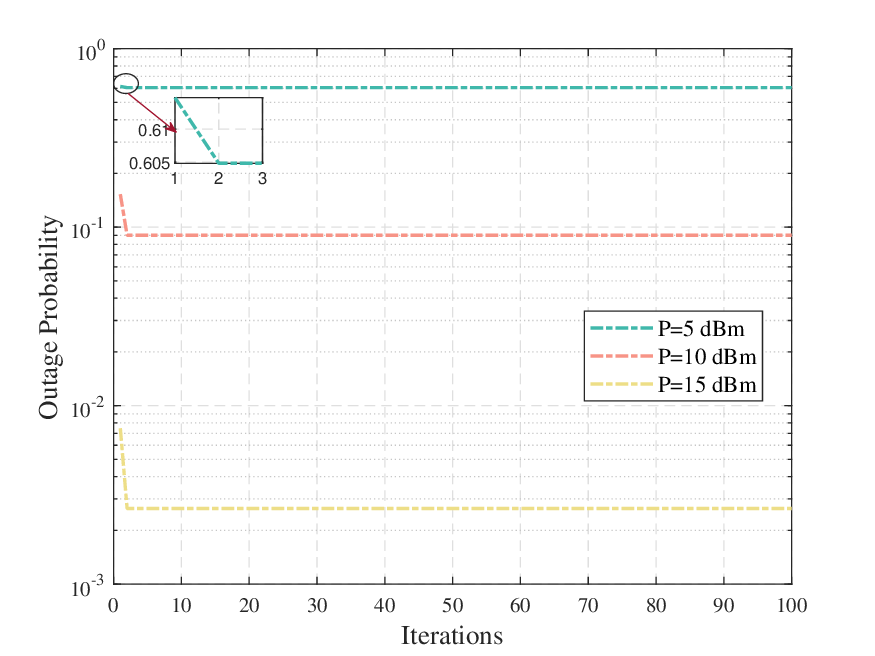}
\caption{Outage probability versus the number of iterations; convergence behavior of our proposed SAC based method for OMA systems, where $N_c=N_e=4$, $W_c=W_e=5$, $R_1=R_2=1$ bps/Hz, $P=5$ dBm, $10$ dBm, and $15$ dBm.}
\end{figure}

In Fig.~5, we investigate the impact of the BS's transmit power, i.e., $P$, on the outage performance of  OMA and NOMA systems. The parameters are set as follows: $N_c=4$, $N_e=6$, $W_c=2$, $W_e=5$ for the general case, and $N_c=N_e=4$, $W_c=W_e=5$ for the special case, with both $R_1$ and $R_2$ equal to $1$ bps/Hz. The term ``Conventional NOMA" refers to the NOMA scheme utilizing a conventional single antenna, analogous to ``Conventional OMA" for the OMA scheme. The ``general case" pertains to scenarios where $N_e\neq N_c$ or $\mu_h\neq \mu_g$, whereas the ``special case" applies when $N_e=N_c$ and $\mu_h=\mu_g$. The phrase ``FAS NOMA, closed-form" describes our proposed NOMA systems' closed-form solution-based scheme, employing an approximation detailed in \eqref{qq36}, contrast to ``FAS NOMA" for our non-approximation-based proposals. Moreover, ``FAS OMA, SCA" signifies our  SCA-based algorithm for OMA systems. The results in Fig.~5 reveal a decrease in the outage probabilities for all schemes as $P$ escalates. Notably, the outage probabilities of FAS-based schemes are consistently lower than those of the ``Conventional NOMA" scheme for $P\geq 6$ dBm, enhancing the  outage probability reduction for both CU and EU, thereby improving fairness. Additionally, the performance disparity between FAS-based schemes and the ``Conventional NOMA" scheme expands with increasing $P$. Moreover, the outage performance of ``FAS NOMA, closed-form, general case" closely aligns with that of the ``FAS NOMA, general case" scheme when $P\geq 20$ dBm, due to the fact that both schemes converge to the optimal solution $\frac{1}{1+\gamma_{th,2}+\gamma_{th,2}/\gamma_{th,1}}$ at $P \geq 20$ dBm.

\begin{figure}
\centering
\includegraphics[width=3.4in]{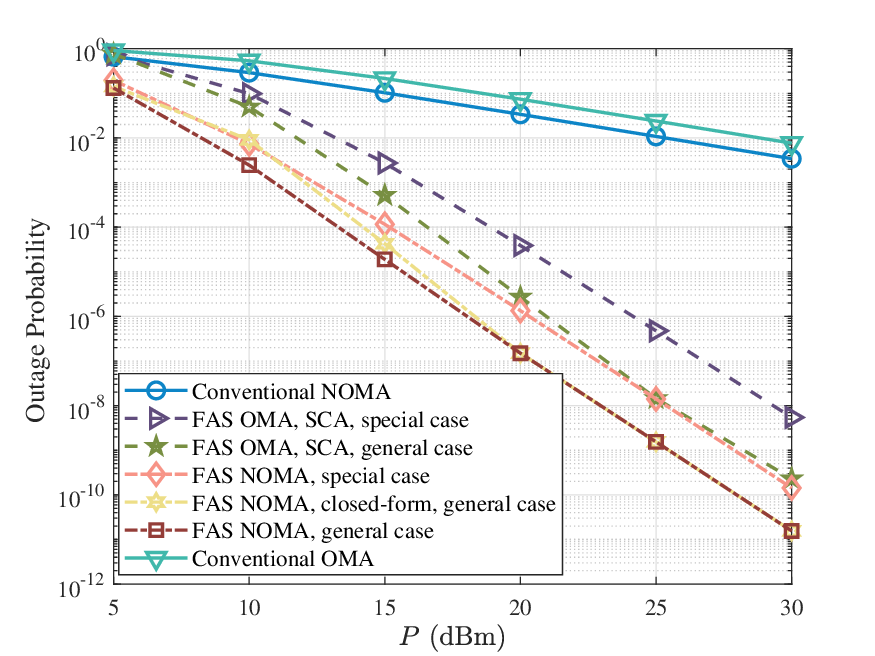}
\caption{Outage probability versus the transmit power of the BS $P$ for OMA and NOMA systems, where $N_c=4$, $N_e=6$, $W_c=2$, $W_e=5$ in general case, $N_c=N_e=4$, $W_c=W_e=5$ in the special case, and $R_1=R_2=1$ bps/Hz.}
\end{figure}

In Fig.~6, we present the effect of the number of the ports at the CU $N_c$ on the outage performance of the OMA and NOMA systems, where $N_e=6$, $W_c=5$, $W_e=1$, $R_1=R_2=1$ bps/Hz, and $P=5$ dBm. From Fig.~6, we can observe that the outage probabilities of all the schemes with FAS decrease as the $N_c$ increases. This is because the more ports at the CU, the better the channel quality of the most favourable port. Therefore, with the increase of $N_c$,  more power allocated to the EU to attain the better fairness between the two users. However, the curve for the ``Conventional NOMA" scheme remains flat since conventional NOMA cannot be able to select the ports. We can also find that the outage performance of the ``FAS OMA, SCA" scheme is better than the ``Conventional NOMA"  scheme when $N_c \geq 13$, which indicates that the influence of the number of ports on the outage probability is significant. Besides, the outage performance of the ``FAS NOMA, closed-form, general case" is close to that of the ``FAS NOMA, general case" when $N_c \geq 8$.

Similar to Fig.~6, we investigate the impact of the number of   ports at the EU $N_e$ on the outage performance of the OMA and NOMA systems in Fig.~7, where $N_c=4$, $W_c=1$, $W_e=5$, $R_1=R_2=1$ bps/Hz, and $P=10$ dBm. It is found from Fig.~7 that the outage probabilities of all the schemes with FAS decrease as $N_e$ increases, since the more ports at the EU, the smaller the outage probability of the EU. The outage performance of the ``FAS NOMA, closed-form, general case" scheme is the same as that of the ``FAS NOMA, general case" scheme when $N_e \geq 12$. This is because the optimal solution is $\frac{1}{1+\gamma_{th,2}+\gamma_{th,2}/\gamma_{th,1}}$ for both schemes when $N_e \geq 12$.

\begin{figure}
\centering
\includegraphics[width=3.4in]{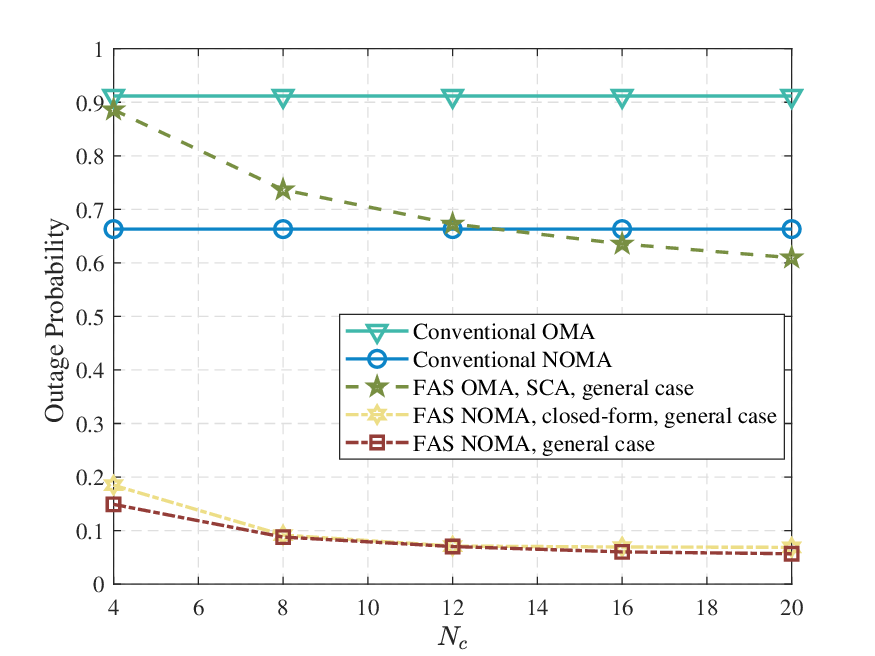}
\caption{Outage probability versus the number of the ports at the CU $N_c$ for the general case in OMA and NOMA systems, where $N_e=6$, $W_c=5$, $W_e=1$, $R_1=R_2=1$ bps/Hz, and $P=5$ dBm.}
\end{figure}

\begin{figure}
\centering
\includegraphics[width=3.4in]{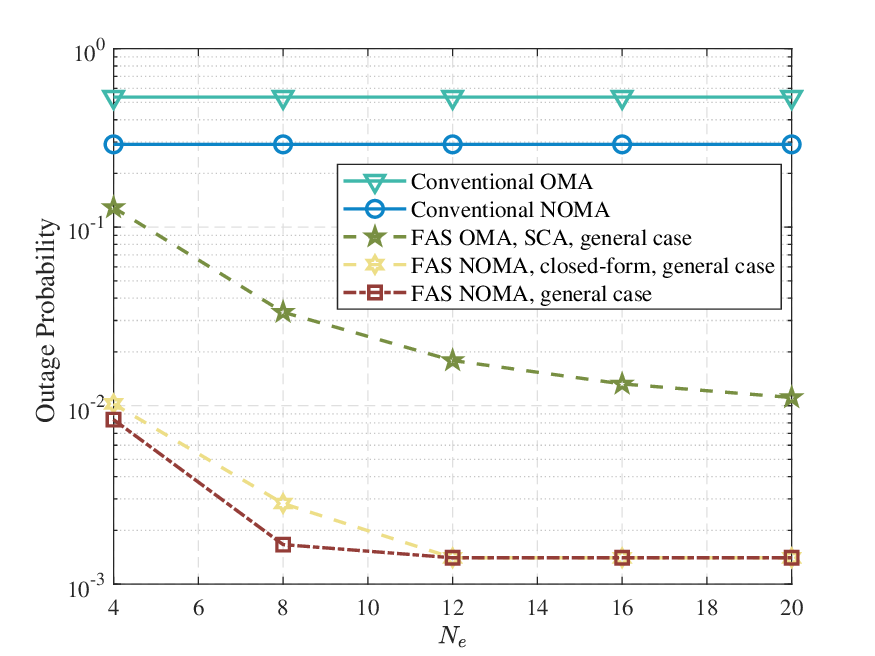}
\caption{Outage probability versus the number of the ports at the EU $N_e$ for the general case in OMA and NOMA systems, where $N_c=4$, $W_c=1$, $W_e=5$, $R_1=R_2=1$ bps/Hz, and $P=10$ dBm.}
\end{figure}

In Fig.~8,  we investigate the effect of the target rate of the CU, denoted by  $R_1$,  on the outage performance of the OMA and NOMA systems, considering $N_c=10$, $N_e=4$ for the general case, and $N_c=N_e=4$ for the special case. The settings include $W_c=W_e=5$, $R_2=1$ bps/Hz, and $P=10$ dBm. The analysis reveals that the outage probabilities for all the schemes increase with $R_1$. This trend is attributed to the increased power allocation to the CU. Although this reduces the outage probability for the CU, it correspondingly increases the outage probability for the EU. A comparative analysis of the ``FAS NOMA, special case" scheme and  ``FAS NOMA, general case" scheme illustrates that an increase in $N_c$ leads to a reduction in outage probability.  The outage performance of the ``FAS NOMA, closed-form, general case" scheme outperforms  the ``FAS NOMA, special case" scheme when $R_1\geq 1.6$ bps/Hz. Furthermore, the performance  gap between the outage performance of the ``FAS NOMA, closed-form, general case" scheme and the ``FAS NOMA, general case" scheme  narrows as $R_1$ increases. This phenomenon can be explained by two factors. Firstly, according to \emph{Lemma 1}, $Q_1\left(\sqrt{\frac{2\mu_h^2}{1-\mu_h^2}}\sqrt{t},\sqrt{\frac{2}{1-\mu_h^2}}\sqrt{\frac{\phi_2}{\sigma_h^2}}\right)$ diminishes with an increase in $R_1$. Secondly, a lower value of $Q_1\left(\sqrt{\frac{2\mu_h^2}{1-\mu_h^2}}\sqrt{t},\sqrt{\frac{2}{1-\mu_h^2}}\sqrt{\frac{\phi_2}{\sigma_h^2}}\right)$ enhances the accuracy of \eqref{qq36}, making $\tilde{\mathbf{P}}^\textrm{out}_{c}(\phi_2)$ more accurate at higher $R_1$ values.

\begin{figure}
\centering
\includegraphics[width=3.4in]{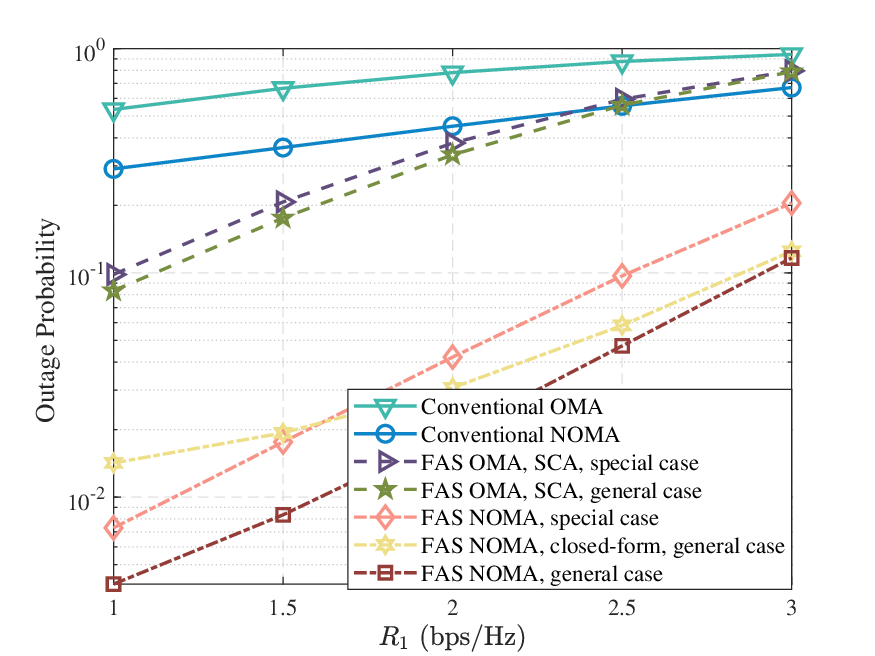}
\caption{Outage probability versus the target rate of the CU $R_1$ for OMA and NOMA systems, where $N_c=6$, $N_e=4$ in the general case, $N_c=N_e=4$ in special case, $W_c=W_e=5$, $R_2=1$ bps/Hz, and $P=10$ dBm.}
\end{figure}

In Fig.~9, we investigate the influence of the EU's target rate, $R_2$, on the outage performance of OMA and NOMA systems, specifying $N_c=4$, $N_e=10$ for the general case, and $N_c=N_e=4$ for the special case, with $W_c=W_e=5$, $R_1=1$ bps/Hz, and $P=10$ dBm. Analysis of Fig.~9 indicates that the outage probabilities for all the schemes employing FAS escalate as $R_2$ increases. A comparative evaluation of the ``FAS OMA, SCA, special case" and ``FAS OMA, SCA, general case" schemes reveal that an increased $N_e$ correlates with reduced outage probabilities. Moreover, the ``FAS NOMA, closed-form, general case" scheme demonstrates superior outage performance compared to the ``FAS NOMA, special case" scheme when $R_2 \geq 1.2$ bps/Hz. Additionally, the discrepancy in outage performance between the ``FAS NOMA, closed-form, general case" and ``FAS NOMA, general case" schemes diminishes as $R_2$ increases, attributed to the enhanced accuracy of $\tilde{\mathbf{P}}^\textrm{out}_e$ with higher $R_2$ values.

\begin{figure}
\centering
\includegraphics[width=3.4in]{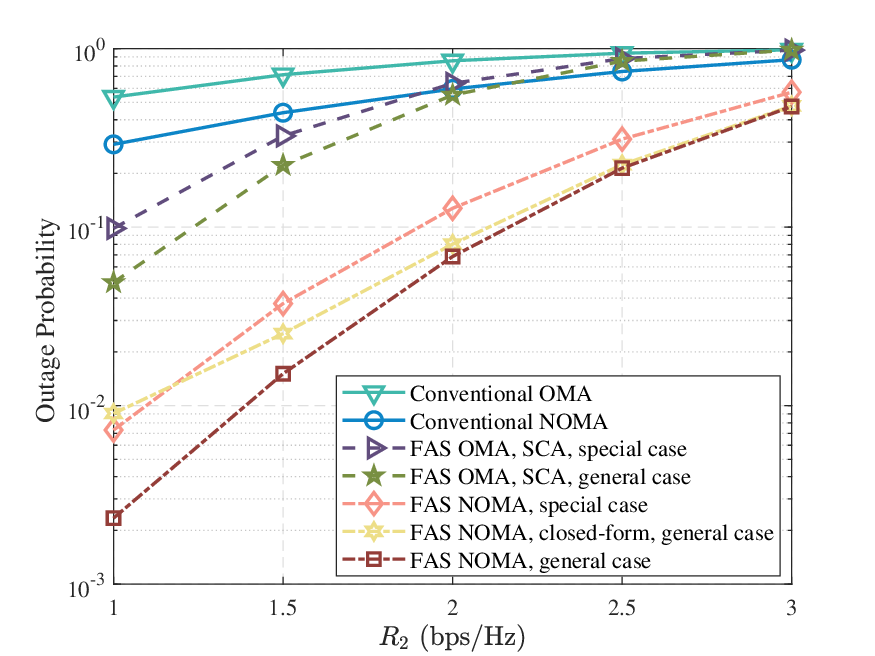}
\caption{Outage probability versus the target rate of the EU $R_2$ for OMA and NOMA systems, where $N_c=4$, $N_e=6$ in the general case, $N_c=N_e=4$ in special case, $W_c=W_e=5$, $R_1=1$ bps/Hz, and $P=10$ dBm.}
\end{figure}

\section{Conclusion}\label{sec6}
In this paper, we considered the downlink NOMA and OMA systems, in which the BS broadcasts the superposition-coded signals to two users and both users are equipped with a single fluid antenna. We addressed the minimization problem of the maximum outage probabilities of the two users for both NOMA and OMA systems. For the NOMA systems, we first considered the special case, i.e., $N_e=N_c$ and $\mu_h=\mu_g$, and obtained the optimal closed-form solution. Then, we obtained the optimal solution by adopting the bisection search for the general case, i.e., $N_e\neq N_c$ or/and $\mu_h\neq \mu_g$. For the OMA systems, we proposed the SCA based algorithm to solve the formulated problem of both the special case and the general case. Numerical results illustrated that our proposed NOMA schemes surpass the conventional NOMA scheme. With FAS, even the performance of OMA can be better than that of the conventional NOMA scheme without FAS.

\appendices
\section{Proof of Lemma 1}
From the Leibniz integral rule
\begin{align}
\frac{\partial}{\partial y}\int_{a(y)}^{b(y)}f(x,y)dx=&\int_{a(y)}^{b(y)}\frac{\partial f(x,y)}{\partial y}dx+f(b(y),y)\frac{\partial b(y)}{\partial y}\nonumber\\ &-f(a(y),y)\frac{\partial a(y)}{\partial y},
\end{align}
we have
\begin{align}\label{aq2}
\frac{\partial\mathbf{P}_c^\textrm{out}(\phi_2)}{\partial \alpha}&=\int_0^\infty e^{-t}\frac{\partial \Upsilon}{\partial \alpha}dt,\\
\label{aq3}\frac{\partial\mathbf{P}_c^\textrm{out}(\phi_1)}{\partial \alpha}&=\int_0^\infty e^{-t}\frac{\partial \Phi}{\partial \alpha}dt,\\
\label{aq4}\frac{\partial\mathbf{P}_e^\textrm{out}}{\partial \alpha}&=\int_0^\infty e^{-t}\frac{\partial \Psi}{\partial \alpha}dt,
\end{align}
where
\begin{align}\label{aq5}
\Upsilon=\left[1-Q_1\left(\sqrt{\frac{2\mu^2}{1-\mu^2}}\sqrt{t},\sqrt{\frac{2}{1-\mu^2}}\sqrt{\frac{\phi_2}{\sigma_h^2}}\right)\right]^{N},\\
\Phi=\left[1-Q_1\left(\sqrt{\frac{2\mu^2}{1-\mu^2}}\sqrt{t},\sqrt{\frac{2}{1-\mu^2}}\sqrt{\frac{\phi_1}{\sigma_h^2}}\right)\right]^{N},\\
\Psi=\left[1-Q_1\left(\sqrt{\frac{2\mu^2}{1-\mu^2}}\sqrt{t},\sqrt{\frac{2}{1-\mu^2}}\sqrt{\frac{\phi_1}{\sigma_g^2}}\right)\right]^{N}.
\end{align}

Taking the first-order partial derivative of $\Upsilon$ with respect to $\alpha$, we have
\begin{align}\label{aq6}
\frac{\partial \Upsilon}{\partial \alpha}=&N\left[1-Q_1\left(\sqrt{\frac{2\mu^2}{1-\mu^2}}\sqrt{t},\sqrt{\frac{2}{1-\mu^2}}\sqrt{\frac{\phi_2}{\sigma_h^2}}\right)\right]^{N-1}\nonumber\\
& \times \frac{\partial \left[1-Q_1\left(\sqrt{\frac{2\mu^2}{1-\mu^2}}\sqrt{t},\sqrt{\frac{2}{1-\mu^2}}\sqrt{\frac{\phi_2}{\sigma_h^2}}\right)\right]}{\partial \alpha}.
\end{align}
Because $1-Q_1\left(\sqrt{\frac{2\mu^2}{1-\mu^2}}\sqrt{t},\sqrt{\frac{2}{1-\mu^2}}\sqrt{\frac{\phi_2}{\sigma_h^2}}\right)$ is a monotonically increasing function with respect to $\frac{\phi_2}{\sigma_h^2}$ \cite{MKSimon20}, and $\phi_2$ in \eqref{q7} is a monotonically decreasing function with respect to $\alpha$, we have $\frac{\partial \left[1-Q_1\left(\sqrt{\frac{2\mu^2}{1-\mu^2}}\sqrt{t},\sqrt{\frac{2}{1-\mu^2}}\sqrt{\frac{\phi_2}{\sigma_h^2}}\right)\right]}{\partial \alpha}<0$. Therefore, we can obtain that $\frac{\partial \Upsilon}{\partial \alpha}<0$, then $\frac{\partial\mathbf{P}_c^\textrm{out}(\phi_2)}{\partial \alpha}<0$.

Similarly, $\frac{\partial\mathbf{P}_c^\textrm{out}(\phi_1)}{\partial \alpha}>0$ and $\frac{\partial\mathbf{P}_e^\textrm{out}}{\partial \alpha}>0$, since $\phi_1$ is a monotonically increasing function with respect to $\alpha$. Thus, $\mathbf{P}_c^\textrm{out}(\phi_2)$ is a decreasing function with respect to $\alpha$, and $\mathbf{P}_c^\textrm{out}(\phi_1)$ and $\mathbf{P}_e^\textrm{out}$ are both increasing with respect to $\alpha$.

\section{Proof of Lemma 2}
Lemma 1 demonstrates that $\mathbf{P}_c^\textrm{out}(\phi_2)$ decreases with respect to $\alpha$, whereas $\mathbf{P}_c^\textrm{out}(\phi_1)$ and $\mathbf{P}_e^\textrm{out}$ both increase as $\alpha$ increases.

Regarding Problem \eqref{q11}, $\mathbf{P}_c^\textrm{out}(\phi_2)=1$ and $0<\mathbf{P}_e^\textrm{out}<1$ when $\alpha=0$. Consider the inequality $\frac{\phi_1}{\sigma_h^2} < \frac{\phi_1}{\sigma_g^2}$. Given this, and noting that the function $1 - Q_1(\alpha, x)$ is monotonically increasing with respect to $x$, where $x$ can represent either $\frac{\phi_1}{\sigma_h^2}$ or $\frac{\phi_1}{\sigma_g^2}$, we can deduce that $\mathbf{P}_c^\textrm{out}(\phi_2)=\mathbf{P}_c^\textrm{out}(\phi_1)<\mathbf{P}_e^\textrm{out}$ for $\alpha=\frac{1}{1+\gamma_{th,2}+\gamma_{th,2}/\gamma_{th,1}}$.  Consequently, an $\hat{\alpha}$ exists such that $\mathbf{P}_c^\textrm{out}(\phi_2)=\mathbf{P}_e^\textrm{out}$ within $0<\alpha\leq \frac{1}{1+\gamma_{th,2}+\gamma_{th,2}/\gamma_{th,1}}$. Additionally, $\max\{\mathbf{P}_c^\textrm{out}(\phi_2),\mathbf{P}_e^\textrm{out}\}=\mathbf{P}_c^\textrm{out}(\phi_2)$ for $0<\alpha\leq \hat{\alpha}$, and $\max\{\mathbf{P}_c^\textrm{out}(\phi_2),\mathbf{P}_e^\textrm{out}\}=\mathbf{P}_e^\textrm{out}$ for $\hat{\alpha}<\alpha\leq \frac{1}{1+\gamma_{th,2}+\gamma_{th,2}/\gamma_{th,1}}$. Therefore, $\max\{\mathbf{P}_c^\textrm{out}(\phi_2),\mathbf{P}_e^\textrm{out}\}$ initially decreases, then increases with $\alpha$, making $\hat{\alpha}$ the optimal solution for Problem \eqref{q11}. Given $\frac{\phi_2}{\sigma_h^2}=\frac{\phi_1}{\sigma_g^2}$, $\hat{\alpha}$ is expressed as
\begin{align}\label{bq1}
\hat{\alpha} = \frac{1}{1+\gamma_{th,2}+\frac{\gamma_{th,2}\sigma_h^2}{\gamma_{th,1}\sigma_g^2}}.
\end{align}

For Problem \eqref{q12}, noting that $\frac{\phi_1}{\sigma_h^2}<\frac{\phi_1}{\sigma_g^2}$ and $1-Q_1\left(\alpha, x \right)$ increases with $x$, $\mathbf{P}_c^\textrm{out}(\phi_1)<\mathbf{P}_e^\textrm{out}$ when $\alpha$ lies between $\frac{1}{1+\gamma_{th,2}+\gamma_{th,2}/\gamma_{th,1}}$ and $\frac{1}{1+\gamma_{th,2}}$. Thus, $\max{\mathbf{P}_c^\textrm{out}(\phi_1),\mathbf{P}_e^\textrm{out}}=\mathbf{P}_e^\textrm{out}$ within this range. The optimum value for Problem \eqref{q12} exceeds $\mathbf{P}_e^\textrm{out}$ when $\alpha = \frac{1}{1+\gamma_{th,2}+\gamma_{th,2}/\gamma_{th,1}}$. As a consequence, the optimal solution for \eqref{q10} is determined to be $\hat{\alpha}$.

\section{Proof of Theorem 1}
After some mathematical manipulation, $\tilde{\mathbf{P}}^\textrm{out}_{c}(\phi_2)=\tilde{\mathbf{P}}^\textrm{out}_{e}$ can be rewritten as
\begin{equation}\label{eq4}
a_1 d_1 \alpha^2+(b_1d_1+c_1-a_1) \alpha-b_1=0,
\end{equation}
where
\begin{align}
\label{eq4-4} a_1=& \ln(N_e/N_c),\\
\label{eq4-1} b_1=& \frac{\gamma_{th,1}\sigma^2}{P\sigma^2_h},\\
\label{eq4-2} c_1=& \frac{\gamma_{th,2}\sigma^2}{P\sigma^2_g},\\
\label{eq4-3} d_1=& 1+\gamma_{th,2}.
\end{align}
When $N_c=N_e$, i.e., $a_1=0$, according to \eqref{eq4}, we can obtain that
\begin{align}\label{q31}
\alpha=&\frac{1}{1+\gamma_{th,2}+\frac{\gamma_{th,2}\sigma_h^2}{\gamma_{th,1}\sigma_g^2}},
\end{align}
When $N_c\neq N_e$, we know that the two possible solutions to \eqref{eq4} are given by
\begin{align}
\label{eq5}\alpha_1=&\frac{-(b_1d_1+c_1-a_1)+\sqrt{(b_1d_1+c_1-a_1)^2+4b_1a_1d_1}}{2a_1d_1},\\
\label{eq6}\alpha_2=&\frac{-(b_1d_1+c_1-a_1)-\sqrt{(b_1d_1+c_1-a_1)^2+4b_1a_1d_1}}{2a_1d_1}.
\end{align}

\section{Proof of Lemma 3}
Taking the first-order partial derivative and the second-order partial derivative of $2^{\frac{R_1}{\beta}}$ with respect to $\beta$, we have
\begin{align}\label{dq1}
\frac{\partial 2^{\frac{R_1}{\beta}}}{\partial \beta}=&-2^{\frac{R_1}{\beta}}\frac{R_1\ln2}{\beta^2}, \\
\frac{\partial^2 2^{\frac{R_1}{\beta}}}{\partial \beta^2}=&2^{\frac{R_1}{\beta}}\frac{R^2_1(\ln2)^2}{\beta^4}+2^{\frac{R_1}{\beta}}\frac{2R_1\ln2}{\beta^3}.
\end{align}
Then taking the first-order partial derivative and the second-order partial derivative of $2^{\frac{R_2}{1-\beta}}$ with respect to $\beta$, we have
\begin{align}\label{dq2}
\frac{\partial 2^{\frac{R_2}{1-\beta}}}{\partial \beta}=&2^{\frac{R_2}{(1-\beta)}}\frac{R_2\ln2}{(1-\beta)^2}, \\
\frac{\partial^2 2^{\frac{R_2}{1-\beta}}}{\partial \beta^2}=&2^{\frac{R_2}{(1-\beta)}}\frac{R^2_2(\ln2)^2}{(1-\beta)^4}+2^{\frac{R_2}{(1-\beta)}}\frac{2R_2\ln2}{(1-\beta)^3}.
\end{align}
Because $\frac{\partial^2 2^{\frac{R_1}{\beta}}}{\partial \beta^2}>0$ and $\frac{\partial^2 2^{\frac{R_2}{1-\beta}}}{\partial \beta^2}>0$, $2^{\frac{R_1}{\beta}}$ and  $2^{\frac{R_2}{1-\beta}}$ are both convex functions with respect to $\beta$.

\end{document}